\documentclass[12pt]{article}
\usepackage[pdftex]{graphicx}
\usepackage{amssymb,amsfonts,amsmath,epsfig,color}
\usepackage{epstopdf}
\makeatletter
\DeclareSymbolFont{AMSb}{U}{msb}{m}{n}
\DeclareSymbolFontAlphabet{\mathbb}{AMSb}
\renewcommand{\section}{\@startsection{section}{1}{\z@}%
                                    {-7ex \@plus -1ex \@minus -.2ex}%
                                    {2.5ex \@plus.2ex}%
                                    {\normalfont\large\scshape\centering}}
\renewcommand{\subsection}{\@startsection{subsection}{2}{\z@}%
                                       {-5ex \@plus -1ex \@minus -.2ex}%
                                       {1.5ex \@plus.2ex}%
                                       {\normalfont\normalsize\scshape}}
\renewcommand{\subsubsection}{\@startsection{subsubsection}{3}{\z@}%
                                       {-5ex \@plus -1ex \@minus -.2ex}%
                                       {1.5ex \@plus.2ex}%
                                       {\normalfont\normalsize\scshape}}

\renewcommand\@seccntformat[1]{\ignorespaces\csname #1name\endcsname\space
                               \csname the#1\endcsname.\quad}   
\newdimen\captionmargin
\newdimen\captionindent
\newdimen\captionwidth
\newcommand{\captionfont}{\slshape}
\newcommand\@captionlabel[1]{\textsc{#1:}\space}
\long\def\@makecaption#1#2{%
  \vskip\abovecaptionskip
  \captionwidth\hsize
  \advance\captionwidth -2\captionmargin
  \sbox\@tempboxa{\@captionlabel{#1}\captionfont #2}%
  \ifdim \wd\@tempboxa >\captionwidth
    \ifdim\captionindent>\z@
      \advance\captionwidth -\captionindent
      \hskip\captionindent
    \fi
    \hskip\captionmargin
    \parbox[t]{\captionwidth}{\leavevmode\hskip-\captionindent
      \@captionlabel{#1}\captionfont #2}%
  \else
    \global \@minipagefalse
    \hb@xt@\hsize{\hfil\box\@tempboxa\hfil}%
  \fi
  \vskip\belowcaptionskip}
\def\eqnarray{%
   \stepcounter{equation}%
   \def\@currentlabel{\p@equation\theequation}%
   \global\@eqnswtrue
   \m@th
   \global\@eqcnt\z@
   \tabskip\@centering
   \let\\\@eqncr
   $$\everycr{}\halign to\displaywidth\bgroup
       \hskip\@centering$\displaystyle\tabskip\z@skip{##}$\@eqnsel
      &\global\@eqcnt\@ne$\;\hfil{##}$\hfil
      &\global\@eqcnt\tw@$\;\displaystyle{##}$\hfil\tabskip\@centering
      &\global\@eqcnt\thr@@ \hb@xt@\z@\bgroup\hss##\egroup
         \tabskip\z@skip
      \cr}
\ifcase \@ptsize
\or
\or
\fi
  \divide\@tempdima\baselineskip
  \@tempcnta=\@tempdima
\makeatother
\begin{document}

%
%

\renewcommand{\theequation}{\arabic{section}.\arabic{equation}}
\renewcommand{\thefigure}{\arabic{figure}}
\newcommand{\gapprox}{%
\mathrel{%
\setbox0=\hbox{$>$}\raise0.6ex\copy0\kern-\wd0\lower0.65ex\hbox{$\sim$}}}
\textwidth 165mm \textheight 220mm \topmargin 0pt \oddsidemargin 2mm
\def\ib{{\bar \imath}}
\def\jb{{\bar \jmath}}

\newcommand{\ft}[2]{{\textstyle\frac{#1}{#2}}}
\newcommand{\be}{\begin{equation}}
\newcommand{\ee}{\end{equation}}
\newcommand{\bea}{\begin{eqnarray}}
\newcommand{\eea}{\end{eqnarray}}
\newcommand{\Identity}{{1\!\rm l}}
\newcommand{\cx}{\overset{\circ}{x}_2}
\def\CN{$\mathcal{N}$}
\def\CH{$\mathcal{H}$}
\def\hg{\hat{g}}
\newcommand{\bref}[1]{(\ref{#1})}
\def\espai{\;\;\;\;\;\;}
\def\zespai{\;\;\;\;}
\def\avall{\vspace{0.5cm}}
\newtheorem{theorem}{Theorem}
\newtheorem{acknowledgement}{Acknowledgment}
\newtheorem{algorithm}{Algorithm}
\newtheorem{axiom}{Axiom}
\newtheorem{case}{Case}
\newtheorem{claim}{Claim}
\newtheorem{conclusion}{Conclusion}
\newtheorem{condition}{Condition}
\newtheorem{conjecture}{Conjecture}
\newtheorem{corollary}{Corollary}
\newtheorem{criterion}{Criterion}
\newtheorem{defi}{Definition}
\newtheorem{example}{Example}
\newtheorem{exercise}{Exercise}
\newtheorem{lemma}{Lemma}
\newtheorem{notation}{Notation}
\newtheorem{problem}{Problem}
\newtheorem{prop}{Proposition}
\newtheorem{rem}{{\it Remark}}
\newtheorem{solution}{Solution}
\newtheorem{summary}{Summary}
\numberwithin{equation}{section}
\newenvironment{pf}[1][Proof]{\noindent{\it {#1.}} }{\ \rule{0.5em}{0.5em}}
\newenvironment{ex}[1][Example]{\noindent{\it {#1.}}}

\thispagestyle{empty}


\begin{center}

{\LARGE\scshape Non-singular quantum improved rotating black holes and their maximal extension
\par}
\vskip15mm

\textsc{R. Torres}
\par\bigskip
{\em
Department of Physics, UPC, Barcelona, Spain.}\\[.1cm]
\vspace{5mm}

\end{center}

\section*{Abstract}
We add a prescription to the Newman-Janis algorithm in order to use it as a means of finding new \emph{extended} (`through $r<0$') rotating black hole spacetimes from static spherically symmetric ones. Then, we apply the procedure to a quantum improved black hole spacetime coming from Quantum Einstein Gravity. The goal is to get a maximally extended spacetime corresponding to a non-singular rotating black hole emulating the standard maximally extended Kerr black hole in regions where quantum effects are negligible. We rigourously check for the existence of scalar curvature singularities in the quantum improved rotating spacetime and we show that it is devoid of  them. We also analyze the horizons and causal structure of the rotating black hole and provide Penrose diagrams for the maximally extended spacetime.

\vskip10mm
\noindent KEYWORDS: Black Holes, Newman-Janis Algorithm, Singularities, Extensions, Asymptotic Safety, Quantum Einstein Gravity.

\vspace{3mm} \vfill{ \hrule width 5.cm \vskip 2.mm {\small
\noindent E-mail: ramon.torres-herrera@upc.edu}}

\newpage
\setcounter{page}{1}

\setcounter{equation}{0}

\section{Introduction}

It is well-known that most astrophysically significant
bodies are rotating. The collapse of a rotating body contributes to the increase of its angular speed while maintaining constant angular momentum. In this way, if the body finally generates a black hole it will be a rotating black hole (RBH). This is the main reason why it is crucial to study RBHs and to analyze their properties.

From a classical point of view, an uncharged (charged) RBH spacetime will be described by a Kerr (Kerr-Newman, resp.) solution. This implies the existence of certain horizons, a specific causal structure and a \emph{singular ring}. However, several authors have suggested that the existence of singularities in the classical solutions has to be considered as a weakness of the theory rather than as a real physical prediction.
Consequently, some have tried to avoid the singularities in the models for RBHs by proposing heuristic regular spacetimes for them (see, for instance, \cite{B&M}\cite{A-A}\cite{LGS}). Other authors, inspired by the work of Bardeen, have taken the path of nonlinear electrodynamics \cite{Bardeen}\cite{A-BI}\cite{A-BII}, which seems to provide the necessary modifications in the energy-momentum tensor in order to avoid singularities in the RBH (see, for instance, \cite{Tosh}\cite{D&G}\cite{Ghosh}).
Yet, another way of addressing the problem of singularities is to take into account that quantum gravity effects should play an important role in the core of black holes, so that it would seem convenient to directly derive the black hole behaviour from an approach to quantum gravity.

In this regard, some regular non-rotating Black Holes inspired in different approaches to Quantum Gravity have appeared in the recent literature (see, for example, \cite{NicolNC}\cite{Frolov2014}\cite{dust2014}\cite{G&P2014}\cite{H&R2014}\cite{M&R} and references therein). For our purposes, let us remark the step in this direction taken by Bonanno and Reuter  in \cite{B&R} by introducing an effective quantum spacetime for spherically symmetric black holes based on the Quantum Einstein Gravity (QEG) approach (see, for instance, \cite{Reuter}\cite{L&R}\cite{R&S}).
The obtained quantum improved Schwarzschild solution indicates
that the horizons and causal structure could be notably modified by quantum corrections and that the BH spacetime could be devoid of singularities.

However, this solution lacks of the rotation that one would expect for realistic black holes. If we want to test quantum improved metrics with astrophysical observations it is necessary to have quantum corrected rotating solutions.
In this line, Reuter and Tuiran \cite{R&T} have tried a direct attack on the problem by using the QEG approach in order to obtain an \textit{improved Kerr solution}. Nevertheless, some problems that had already appeared in the non-rotating case \cite{B&R} become now much more important. Namely, in the QEG approach and through the use of the Functional Renormalization Group Equation, first, one finds the running Newton constant $G(k)$
depending on the considered energy scale $k$ \cite{Reuter}
\begin{equation}\label{Gk}
G(k)=\frac{G_0}{1+\omega G_0 k^2},
\end{equation}
where $\omega$ is a constant and $G_0$ is the standard gravitational constant.
Then, one converts the energy scale dependence into a position dependence, what can be written as
\begin{equation}\label{kP}
k(P)=\frac{\xi}{d(P)},
\end{equation}
where $\xi$ is a constant (to be fixed) and $d(P)$ is the \textit{distance scale} that provides the relevant cutoff when a test particle is located at a point $P$.
If the distance scale must be diffeomorphism invariant then one could write
\[
d(P)=\int_{\mathcal C} \sqrt{|ds^2|},
\]
where $\mathcal C$ is a curve from a reference point $P_0$ to $P$. The problem is that there is a great deal of freedom in choosing $\mathcal C$ for the RBH case and that there is not a unique natural choice for the distance scale. Nevertheless, If we write the RBH spacetime in Boyer-Lindquist-like coordinates $(t,r,\theta,\phi)$, one can restrict the dependence of $d$ on the coordinates simply by taking into account that for a \emph{stationary} and \emph{axially symmetric} spacetime $d=d(r, \theta)$ only (so that $G=G(r,\theta)$). In \cite{R&T} it is argued that the dependence of $d$ on $\theta$ should be asymptotically subdominant (i.e., negligible for $r\rightarrow \infty$). It is also argued that the dependence on $\theta$ should not be too important for $r$ of the order of Planck length. However, a specific expression for the angular dependence was not found.

Our aim in this article is to obtain a quantum improved rotating black hole by using as alternative approach the Newman-Janis (NJ) algorithm \cite{N&J}, which allows to get a rotating solution from a static spherically symmetric one.
The use of the standard NJ algorithm with the goal of obtaining \emph{non-singular} black hole solutions was suggested by Bambi and Modesto in \cite{B&M}. Here, we will see that, in general, the strict standard approach consisting of five steps \cite{Drake&Szek}\cite{B&M}
must be supplemented with an extra prescription if we want to get a well-behaved \emph{extended} (`through $r<0$') RBH spacetime from the algorithm.

Equipped with this prescription we will apply it to the aforementioned quantum improved Schwarzschild solution \cite{B&R}. The goal is to find a regular rotating black hole spacetime emulating the standard maximally extended Kerr solution in regions where quantum effects are negligible. Once the correct maximally extended improved RBH spacetime is found we will analyze its properties.
In particular we will be interested in
rigourously proving the absence of scalar curvature singularities by studying the complete set of algebraically independent curvature scalars.
We will also analyze the horizons and causal structure of the improved spacetime which will be compared with those of the classical Kerr solution.

The article has been divided as follows. Section \ref{NJA} is devoted to the Newman-Janis algorithm and the enhancements required to provide us with correct extended (`through $r<0$') spacetimes for rotating black holes. In section \ref{secISS} the quantum  improved Schwarzschild solution is introduced and the enhanced N-J algorithm is used to provide and extended rotating black hole. The regularity of the obtained spacetime is shown in section \ref{secReg}, while section \ref{secEC} is devoted to the study of the fulfillment of the energy conditions. The possible global structure of the BH spacetime depending on both its mass and its angular momentum is studied in section \ref{secGE}.
Finally, the results are discussed in section \ref{secCon}.

\section{Newman-Janis algorithm and maximally extended spacetimes}\label{NJA}

The standard Newman-Janis algorithm is a five-step procedure for generating new solutions of Einstein's equations by using as a seed solution a static spherically symmetric one \cite{N&J}\cite{Drake&Szek}. The seed solution can always be written as
\begin{equation}\label{seed}
ds^2=-f(r) dt^2+g(r) dr^2+r^2 d\Omega^2.
\end{equation}
The five steps are \cite{Drake&Szek}:
\begin{enumerate}
\item Rewrite the seed line element in advanced null coordinates.
\item Express the contravariant form of the metric in terms of a null tetrad $Z^\mu_a$.
\item Extend the coordinates $x^\rho$ to a new set of complex coordinates
\[
x^\rho \rightarrow \tilde{x}^\rho=x^\rho+i y^\rho(x^\sigma)
\]
and let the null tetrad vectors $Z^\mu_a$ undergo a transformation
\[
Z^\mu_a \rightarrow \tilde{Z}^\mu_a(\tilde{x}^\rho, \bar{\tilde{x}}^\rho).
\]
Require that the transformation recovers the old tetrad and metric when $\tilde{x}^\rho=\bar{\tilde{x}}^\rho$.
\item Obtain a new metric by making a complex coordinate transformation
\[
\tilde{x}^\rho=x^\rho+i \gamma^\rho(x^\sigma)
\]
\item Apply a coordinate transformation $u=t+F(r)$, $\phi=\varphi+H(r)$ to transform the metric to Boyer-Lindquist-type coordinates.
\end{enumerate}

As it is well-known this procedure has been successfully applied to the Schwarzschild solution in order to get (in a straightforward way) Kerr's solution. It is convenient to remark here that, as Kerr's solution reveals, even if the final rotating solution of this general procedure can be singular this does not mean that $r=0$ cannot be traversed. On the contrary, consider Kerr's case in which $r=0$ represents a whole disk \cite{B&L}\cite{H&E}. Only the boundary of the disk ($r=0$, $\theta=\pi/2$) is singular so that $r=0$ can be \textit{traversed} through $\theta\neq \pi/2$. In this way, the disk can be considered as a two-sided aperture to a second sheet on which $r$ is negative, what provide us with an analytic extension of the solution.
In Kerr's solution the standard NJ algorithm can be applied directly obtaining a natural extension for $r<0$ because the seed Schwarzschild solution has $f(r)=1-2 m/r=g^{-1}(r)$. In this way, when one considers negative values for $r$ this is clearly mathematically feasible and physically equivalent to deal with the geometry generated by a negative mass. As a consequence, when the function $f$ undergoes the process of complexification becoming $\bar f=\bar f(r,\theta)$ this eventually provide us with a RBH spacetime that is well-behaved for all $r\in \Re$. In other words, we directly get a natural extension through $r<0$ because, previously, the seed spacetime covered with positive values of $r$ and the seed spacetime covered with negative values of $r$, were both well-behaved.

Now, in general, when one uses the NJ-algorithm one would like to extend the new found solution (even more, if the considered solution is regular) \textit{beyond} $r=0$. However, if one insists in emulating the `$r<0$ extension' used in the standard Kerr solution, one has to take into account that for $r<0$ both $f(r)$ and $g(r)$ could have problems from a mathematical point of view (for example, do they exist and are real?) and from a physical point of view (is the found solution meaningful for $r<0$?). In this way, if one wants the NJ-algorithm to provide a natural extended solution through $r<0$ one needs to add a preliminary prescription:

\vspace{0.2 cm}
\textsl{Deduce, if possible, the correct behaviour (both from a mathematical and a physical point of view) of $f(r)$ and $g(r)$ in the spacetime covered with negative values of $r$.}
\vspace{0.2 cm}

In practice, this often requires rethinking the method\footnote{Note that \textit{the method} usually will not imply the use of GR.} used to reach the original seed (\ref{seed}), but now considering that $r<0$, what could be nontrivial in most cases.
In order to exemplify this, let us now find an extended (through $r<0$) non-singular quantum improved solution for a rotating black hole spacetime.

\section{Improved rotating solution}\label{secISS}

The \textit{renormalization group improved} Schwarzschild solution found by Bonanno and Reuter \cite{B&R} can be written as
\begin{equation}\label{RGISch}
ds^2=-f(r) dt^2+f(r)^{-1} dr^2+ r^2 d\Omega^2.
\end{equation}
where
\[
f(r)=1-\frac{2 G(r) m}{r}
\]
and
\begin{equation}
G(r)=\frac{G_0 r^3}{r^3+\tilde{\omega} G_0 (r+\gamma G_0 m)}, \label{GR}
\end{equation}
$G_0$ is Newton's universal gravitational constant, $m$ is the mass measured by an observer at infinity and $\tilde{\omega}$ and $\gamma$ are constants coming from the non-perturbative renormalization group theory and from an appropriate ``cutoff identification", respectively.
The preferred theoretical value of $\gamma$ is $\gamma=9/2$
while
it can be deduced that the precise value of $\tilde \omega$ is $\tilde \omega=167/30\pi$. In fact, the properties of the solution do not rely on their precise values as long as they are strictly positive. A relevant fact with regard to $\tilde \omega$ is that it carries the quantum modifications. In effect, if we make explicit Planck's constant in (\ref{GR}), one gets $\tilde \omega =167 \hbar/30\pi $ and, thus, $\tilde \omega=0$ would turn off the quantum corrections.

Now, in order to see the problems with the standard NJ-algorithm \cite{Drake&Szek}\cite{B&M}, we can try to blindly apply it to the solution (\ref{RGISch}) in order to get a quantum improved RBH spacetime. The five steps for this case would be:
\begin{enumerate}
\item
The coordinate change $du=dt-dr/f(r)$ allows us to write the metric in advanced null coordinates as
\[
ds^2=-f(r) du^2-2 du dr + h(r) d\Omega^2,
\]
where $h(r)=r^2$.
\item
The null tetrad $Z^\mu_a=(l^\mu,n^\mu,m^\mu,\bar m^\mu)$ satisfying
$l_\mu n^\mu=-m_\mu\bar m^\mu=-1$ and $l_\mu m^\mu=n_\mu m^\mu=0$
can be chosen as
\[
l^\mu=\delta^\mu_r,\hspace{1 cm} n^\mu=\delta^\mu_u-\frac{f(r)}{2}\, \delta^\mu_r,\hspace{1 cm}
m^\mu=\frac{1}{\sqrt{2 h(r)}} \left(\delta^\mu_\theta+\frac{i}{\sin\theta} \delta^\mu_\phi\right)
\]
so that $g^{\mu\nu}=-l^\mu n^\nu-l^\nu n^\mu+m^\mu \bar m^\nu+m^\nu \bar m^\mu$.
\item We perform the standard coordinate change
\[
r'=r+i\, a \cos\theta, \hspace{1 cm}  u'=u-i\, a \cos\theta.
\]
and demand $r'$ and $u'$ to be real.
In this way the null tetrad transforms into ($Z'^\mu_a=Z^\nu_a \partial x^{\mu'}/\partial x^\nu$)
\[
l'^\mu=\delta^\mu_r, \hspace{.5cm} n'^\mu=\delta^\mu_u-\frac{\bar f(r')}{2}\, \delta^\mu_r,\hspace{.5 cm}
m'^\mu=\frac{1}{\sqrt{2 \bar{h}(r')}} \left(\delta^\mu_\theta+\frac{i}{\sin\theta} \delta^\mu_\phi+i\, a \sin\theta (\delta^\mu_u-\delta^\mu_r)\right)
\]
The functions $\bar f$ and $\bar h$ come from the complexification of $f$ and $h$ and, for the moment, we only know that they must be real and that they must reproduce Kerr solution if the quantum effects are turned off ($\tilde \omega=0$). This is possible if the functions are chosen in the usual manner \cite{N&J}\cite{Drake&Szek}\cite{B&M}, i.e., by using the complexification
\[
\frac{1}{r}\rightarrow \frac{1}{2} \left(\frac{1}{r'}+\frac{1}{\bar r'} \right),\ \ r^2\rightarrow r' \bar r'
\]
that provide us with
\begin{eqnarray}
\bar h&=& r^2+ a^2 \cos^2\theta=\Sigma\nonumber\\
\bar f&=&1- \frac{2 \bar G m r}{\Sigma},\label{barf}
\end{eqnarray}
where there is still some freedom in choosing the function $\bar G$.

\item
The new non-zero metric coefficients can be computed to be
\begin{eqnarray}\label{metcoef}
g_{uu}&=&-\bar f(r,\theta), \hspace{.5cm} g_{ur}=-1, \hspace{.5cm} g_{u\phi}=-a \sin^2\theta [1-\bar f (r,\theta)]\\
g_{r\phi}&=&a \sin^2 \theta, \hspace{.5cm} g_{\theta\theta}=\Sigma, \hspace{.5cm} g_{\phi\phi}=\sin^2\theta [\Sigma +a^2 \sin^2\theta (2-\bar f)]\nonumber
\end{eqnarray}
\item In order to get the metric in Boyer-Lindquist type coordinates we perform the coordinate change
$u=t+F(r)$, $\phi=\varphi+H(r)$
where
\begin{equation}\label{transBL}
F(r)=\frac{r^2+a^2}{\bar f(r,\theta) \Sigma +a^2\sin^2 \theta},\hspace{.5cm}
H(r)=\frac{a}{\bar f(r,\theta) \Sigma+a^2\sin^2 \theta}
\end{equation}
and $\bar f$ and $\bar h$ are such that $F$ and $H$ must be functions of $r$ alone.
In principle, one could conceive a general $\bar G$ (thus, $\bar f$) with the form
\[
\bar G (r,\theta;\alpha,\beta,\delta)=G_0\frac{r^{3+\alpha} \Sigma^{-\alpha/2}}{r^{3+\alpha} \Sigma^{-\alpha/2}+\tilde\omega G_0 (r^{1+\beta }\Sigma^{-\beta/2}+\gamma G_0 m r^\delta \Sigma^{-\delta/2})}
\]
where $\alpha$, $\beta$ and $\delta$ are parameters. However,
note that (\ref{transBL}) implies
\[
\Sigma \bar f(r,\theta)+a^2 \sin^2\theta=D(r),
\]
and substituting $\bar f $ using (\ref{barf}) one immediately sees that $\bar G=\bar G(r)$. In other words, $\bar G$ cannot depend on $\theta$. Thus, $\alpha=\beta=\delta=0$ and one is left with
the straightforward case in which
\begin{equation}\label{Grot}
\bar G(r)=G(r)=\frac{G_0 r^3}{r^3+\tilde{\omega} G_0 (r+\gamma G_0 M)}.
\end{equation}
In effect, in this case $F$ and $H$ are really functions of $r$ alone since
\[
F(r)=\frac{r^2+a^2}{r^2+a^2-2 m \bar G(r) r} \ \  \mbox{and}\ \
H(r)=\frac{a}{r^2+a^2-2 m \bar G(r) r}
\]
and, therefore, it is possible to write the solution in Boyern-Lindquist type coordinates.

Let us stop here the algorithm in order to consider what would happen to the metric coefficients (\ref{metcoef}) if, following Kerr's example, one tries to analytically extend the solution through $r=0$ by just considering negative values for $r$.
Clearly, one gets that this extension is not admissible from a physical point of view. It suffices to consider (\ref{Gk}) in which $G(k)\geq 0$ (and $G(k) \stackrel{k\rightarrow \infty}{\rightarrow} 0$) and compare it with (\ref{Grot}) that takes negative values around $r=0$ for negative values of $r$. This is due to the fact that when one goes from (\ref{Gk}) to (\ref{GR}) \cite{B&R} one assumes that $r$ is non-negative. Therefore, as argued in the previous section, we should have first computed the correct behaviour of the improved Schwarzschild solution for negative values of $r$ (what requires rethinking the derivation of the improved solution). We have done this in the appendix obtaining $G(r)$ for all $r\in \Re$ as
\begin{equation}\label{Gallr}
G(r)=\frac{G_0 |r|^3}{|r|^3+\tilde\omega G_0 (|r|+\gamma G_0 m)}.
\end{equation}
This running $G$ is a non-negative, even and $C^2$ function (see fig.\ref{fGallr}).
\begin{figure}
\includegraphics[scale=1]{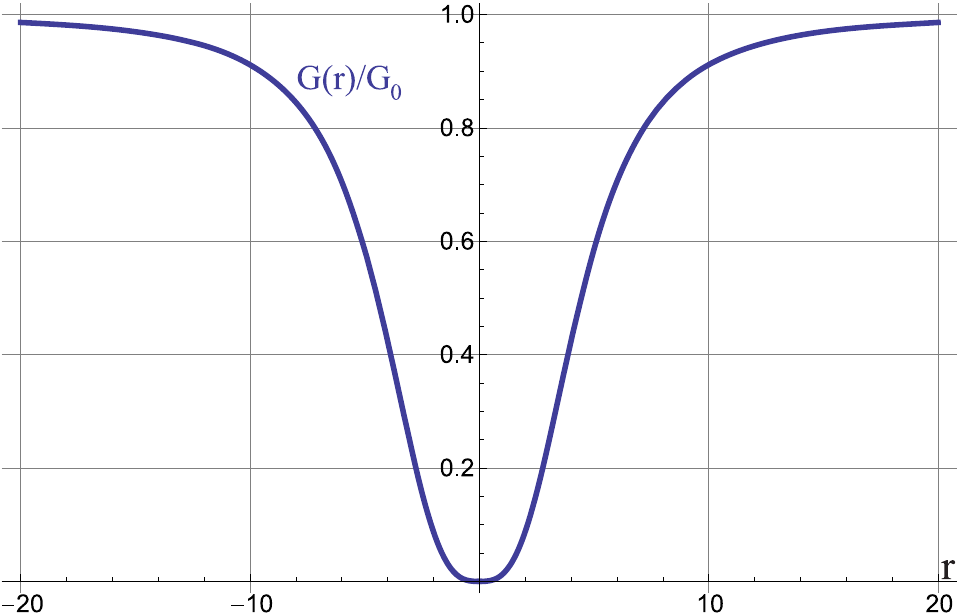}
\caption{\label{fGallr} A plot of $G(r)/G_0$ for a BH mass $m=$10 Planck masses.}
\end{figure}
Only if one applies the N-J algorithm using this \textit{running} $G$ (and, thus, defining sensible functions $f(r)$ and $g(r)$ in the seed spacetime covered with $r<0$ in (\ref{seed})) one can obtain a physically meaningful naturally extended RBH spacetime.

From here, the JN-algorithm tell us that
the new non-zero metric coefficients will be
\begin{eqnarray*}
g_{tt}&=&-\left(1-\frac{2 G(r) m r }{\Sigma} \right), \hspace{.5cm} g_{t\varphi}=-\frac{2  G(r) m r}{\Sigma}a \sin^2\theta, \hspace{.5cm}
g_{rr}=\frac{\Sigma}{\Delta_{\tilde\omega}}\\
g_{\theta\theta}&=&\Sigma, \hspace{.5cm} g_{\varphi\varphi}=\sin^2\theta \left(r^2+a^2+\frac{2 G(r) m r}{\Sigma} a^2 \sin^2\theta\right)
\end{eqnarray*}
where
\[
\Delta_{\tilde\omega}\equiv r^2+a^2-2 G(r) m r.
\]
and $G(r)$ is defined in (\ref{Gallr}).
%
%
%
Thus, the line element can be written in the familiar Boyer-Lindquist form as
\begin{equation}\label{metRBH}
ds^2=-\frac{\Delta_{\tilde\omega}}{\Sigma} (dt-a \sin^2\theta d\phi)^2+\frac{\Sigma}{\Delta_{\tilde\omega}} dr^2+
\Sigma d\theta^2+\frac{\sin^2\theta}{\Sigma} (a dt-(r^2+a^2) d\phi)^2,
\end{equation}
where the quantum corrections are all included in $\Delta_{\tilde\omega}$ (which explains why we have chosen the subindex $\tilde\omega$).
\end{enumerate}

\section{Regularity}\label{secReg}

In order to this spacetime to be devoid of scalar curvature singularities one should proof that all the algebraically independent second order curvature scalars in this spacetime are finite. While the metric (\ref{metRBH}) is singular at $\Delta_{\tilde\omega}=0$ and at $\Sigma=0$ it is easy to check (see \cite{TorresGRG} for the general case) that $\Delta_{\tilde\omega}=0$ is just a coordinate singularity and that it defines horizons in the spacetime (which will be analyzed later). With regard to $\Sigma=0$, the regularity checking is more involved. On the one hand, it is easy to see \cite{TorresGRG} that this spacetime is Petrov type D and Segre type [(1, 1) (1 1)]. This implies that the spacetime has only six real algebraically independent second order curvature scalars \cite{ZM} that are collected in $\{\mathcal R,I,I_6,K\}$,
where $\mathcal R$ is the curvature scalar and the rest of the invariants are defined as\footnote{Here the invariants are written in tensorial form. See \cite{ZM} for their spinorial form.}
\begin{eqnarray*}
I_6&\equiv&
\frac{1}{12}
{S_\alpha}^\beta {S_\beta}^\alpha,\\
I &\equiv&
\frac{1}{24}
\bar{C}_{\alpha\beta\gamma\delta}
\bar{C}^{\alpha\beta\gamma\delta},\\
K &\equiv&
\frac{1}{4}
\bar{C}_{\alpha\gamma\delta\beta}
S^{\gamma\delta} S^{\alpha\beta},
\end{eqnarray*}
where ${S_\alpha}^\beta \equiv {R_\alpha}^\beta-
{\delta_\alpha}^\beta \mathcal{R}/4$ and
$\bar{C}_{\alpha\beta\gamma\delta}\equiv
(C_{\alpha\beta\gamma\delta} +i\ *C_{\alpha\beta\gamma\delta})/2$
is the complex conjugate of the selfdual Weyl tensor
being $*C_{\alpha\beta\gamma\delta}\equiv
\epsilon_{\alpha\beta\mu\nu} C^{\mu\nu}_{\ \ \gamma\delta}/2$ the dual of the Weyl tensor
\footnote{Note that $\mathcal R$ and $I_6$ are real, while $I$ and $K$ are complex. Therefore, there are, indeed, only 6 independent real scalars.}.

The fact that $G(r)$ is not a $C^3$ function indicates that we cannot directly apply the general result of regularity in \cite{TorresGRG}. However, the proof of regularity can be carried out in similar terms.
By computing the curvature scalar $\mathcal{R}$ for our BH one finds
\[
\mathcal{R}=\frac{2 m (2 G' + r G'')}{\Sigma}.
\]
In order to see that this is finite along any path approaching $(r=0,\theta=\pi/2)$,
let us now define the dimensionless quantity $\xi\equiv a \cos\theta/r$ and $\xi^*$, its value in the limit along a chosen path approaching $r=0$. Taking into account that $G'(0)=G''(0)=0$, one finds
\begin{align*}
\mathcal R &\rightarrow& &\frac{4 m G'''(0^+)}{1+\xi^{*2}}& \hspace{1cm} &\mbox{ if } \xi^* \mbox{ finite and we approach $r=0$ from positive values of $r$, }&\\
\mathcal R &\rightarrow& &\frac{4 m G'''(0^-)}{1+\xi^{*2}}& \hspace{1cm} &\mbox{ if } \xi^* \mbox{ finite and we approach $r=0$ from negative values of $r$, }&\\
\mathcal R &\rightarrow& &0&  \hspace{1cm} &\mbox{ if } \xi^* \mbox{ infinite}.&
\end{align*}
Since $G'''(0^+)=-G'''(0^-)=6/(\gamma \tilde\omega G_0 m)$, $\mathcal R$ would be finite along any path.

On the other hand,
\begin{align*}
I_6 &\rightarrow& &\frac{\{m G'''(0^+)\}^2 \xi^{*4}}{3 (1+\xi^{*2})^4}&\hspace{1cm} &\mbox{ if } \xi^* \mbox{ finite and we approach $r=0$ from positive values of $r$,  }&\\
I_6 &\rightarrow& &\frac{\{m G'''(0^-)\}^2 \xi^{*4}}{3 (1+\xi^{*2})^4}& \hspace{1cm} &\mbox{ if } \xi^* \mbox{ finite and we approach $r=0$ from negative values of $r$,  }&\\
I_6 &\rightarrow& &0&\hspace{1cm} &\mbox{ if } \xi^* \mbox{ infinite},&
\end{align*}
what again is finite along any path.

With regard to $I$,
\begin{align*}
I &\rightarrow& &\frac{\{m G'''(0^+)\}^2 \xi^{*4}}{9 (1-i \xi^*)^4 (1+\xi^{*2})^2 }&  \hspace{1cm} &\mbox{ if } \xi^* \mbox{ finite and we approach $r=0$ from positive values of $r$,  }&\\
I &\rightarrow& &\frac{\{m G'''(0^-)\}^2 \xi^{*4}}{9 (1-i \xi^*)^4 (1+\xi^{*2})^2 }&  \hspace{1cm} &\mbox{ if } \xi^* \mbox{ finite and we approach $r=0$ from negative values of $r$,  }&\\
I &\rightarrow& &0& \hspace{1cm} &\mbox{ if } \xi^* \mbox{ infinite},&
\end{align*}
so that $I$ is finite along any path reaching $r=0$.

Finally, we get
\begin{align*}
K &\rightarrow& &\frac{2 \{ m G'''(0^+)\}^3 \xi^{*6}}{3 (1-i \xi^*)^2 (1+\xi^{*2})^5 }& \hspace{1cm} &\mbox{ if } \xi^* \mbox{ finite and we approach $r=0$ from positive values of $r$, }&\\
K &\rightarrow& &\frac{2 \{ m G'''(0^-)\}^3 \xi^{*6}}{3 (1-i \xi^*)^2 (1+\xi^{*2})^5 }& \hspace{1cm} &\mbox{ if } \xi^* \mbox{ finite and we approach $r=0$ from negative values of $r$, }&\\
K &\rightarrow& &0& \hspace{1cm} &\mbox{ if } \xi^* \mbox{ infinite},&
\end{align*}
what is finite along any path reaching $r=0$.

Therefore, we conclude that there are not scalar curvature singularities in the spacetime.

\section{Effective energy-momentum and energy conditions}\label{secEC}

The spacetime metric (\ref{metRBH}) has not been obtain by using Einstein's equations. However, it is still possible to consider an \textit{effective energy-momentum tensor} defined through
\[
8 \pi G_0 T_{\mu\nu}\equiv R_{\mu\nu}-\frac{1}{2} \mathcal R g_{\mu\nu}.
\]
For this spacetime it is easy to show that the effective energy-momentum tensor is type I \cite{H&E}
with
\begin{eqnarray*}\label{mupr}
\mu=-p_\bot&=&\frac{m r^2  G'}{4\pi G_0 \Sigma^2}\\
p_\|&=&- \frac{2 a^2 \cos^2{\theta} G'+r \Sigma G''}{8\pi G_0\Sigma^2} m.
\end{eqnarray*}
where $\mu$, $p_\bot$ and $p_\|$ are the (effective) vacuum energy density, \textit{radial} and tangential pressures, respectively, in the orthonormal basis in which $T$ diagonalizes.

\begin{figure}[t]
\includegraphics[scale=.7]{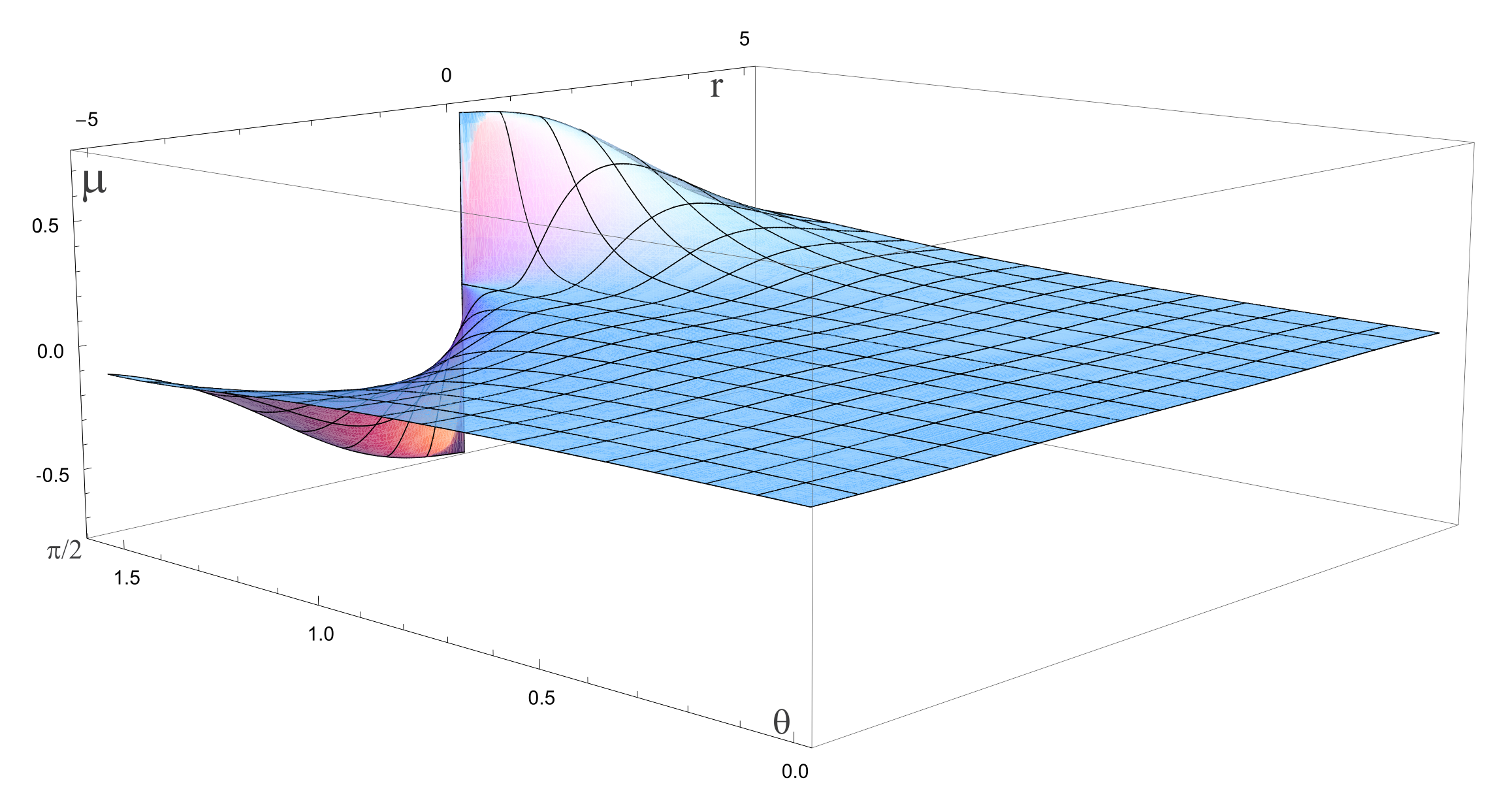}
\caption{\label{fdensity} A plot of the effective vacuum energy-density $\mu$ for a black hole with $m=$10 Planck masses and $a=7$. Note that $\mu$ concentrates around $\theta=\pi/2$ with a maximum at $r=0$. As explained, Kerr's \textit{singular ring} is replaced by a \textit{regular belt}.   }
\end{figure}

The weak energy conditions require
\[
\mu\geq 0,\hspace{1cm}\mu+p_\parallel\geq 0,\hspace{0.5cm} \mbox{and}\hspace{0.5cm} \mu+p_\perp\geq 0.
\]
This is violated for $r<0$ since the effective vacuum density satisfies $\mu< 0$ in this asymptotically flat region ($G'<0$ there -see fig.\ref{fGallr}). On the other hand, $\mu>0$ for $r>0$ and $\mu=0$ for $r=0$ \emph{and} $\theta\neq \pi/2$. In this way, an \textit{observer} can cross $r=0$ with $\theta\neq\pi/2$ measuring an effective energy-density that varies continuously from positive to negative values or viceversa. However, the value of $\mu$ reaches its absolute maximum value when approaching $r=0$, $\theta=\pi/2$ which is
$|\mu|=3/(4\pi \gamma \tilde\omega G_0^2)$. This is of the order of the Planck energy density, i.e., around $10^{113}\ J/m^3$ (in the International System of Units). A plot of the the effective vacuum energy density around $r=0$ is shown in figure \ref{fdensity}.

The spacetime also violates the weak energy conditions in the region with $r>0$ around $r=0$. Specifically, the inequality that is not satisfied is $\mu+p_\parallel\geq 0$.
In order to check this it suffices to consider its expression around $r=0$:
\[
\mu+p_\parallel=-\frac{3  \sec^2\theta}{4 a^2 G_0^2 \gamma \tilde\omega} r^2+\mathcal O(r^3),
\]
which satisfies $\mu+p_\parallel<0$ for $r>0$.

This is not surprising since the absence of singularities implies that the spacetime should violate at least one of the conditions appearing in the standard singularity theorems. We are just showing that the usual energy conditions appearing in the standard singularity theorems are violated.

\section{Global structure}\label{secGE}

As stated in section \ref{secReg}, there is a coordinate singularity at $\Delta_{\tilde\omega}=0$. As usual \cite{B&L}, it is possible to extend the coordinate system beyond $\Delta_{\tilde\omega}=0$ using the coordinate change in section \ref{secISS} (first step in the NJ algorithm) with straightforward predictable consequences. The coordinate $r$ changes its character from spacelike when $\Delta_{\tilde\omega}>0$ to timelike when $\Delta_{\tilde\omega}<0$. Therefore, the boundaries $\Delta_{\tilde\omega}=0$ between these regions are horizons of the spacetime. Classically ($\tilde\omega=0$), there are two solutions to $\Delta\equiv\Delta_{\tilde\omega=0}=0$:
\[
r_\pm=G_o m\pm\sqrt{G_0^2 m^2-a^2},
\]
corresponding to an inner $r_-$ and an outer $r_+$ (\textit{Cauchy} and \textit{event}, respectively) horizons.
Now, in order to get the quantum corrected horizons we should solve
\[
\Delta_{\tilde\omega}=r^2+a^2-2 G(r) m r=0,
\]
which is equivalent to finding the roots of a fifth-degree polynomial. Even if there is not a general formula for the roots in this case we can analyze the general behaviour of the horizons by taking into account the following
\begin{itemize}
\item Since $G(r)\geq 0$ there will not be roots for negative values of $r$. I.e., there are no horizons in the $r<0$ asymptotically flat regions.
\item At large distances, $G\sim G_0$ so that one recovers the behaviour for the Kerr solution. In particular, $\Delta_{\tilde\omega}>0$ and $r$ will be a spacelike coordinate.
\item For $r\simeq 0$ ($a\neq 0$) we have $\Delta_{\tilde\omega}>0$ thanks to the effect of the rotation and, again, $r$ will be a spacelike coordinate. Note that this is what happened in the classical Kerr solution, however now the inner region is in full quantum regime ($G\sim 0$). We illustrate the differences between the classical and the quantum case in fig.\ref{fDC&Q}. %
\begin{figure}
\includegraphics[scale=1]{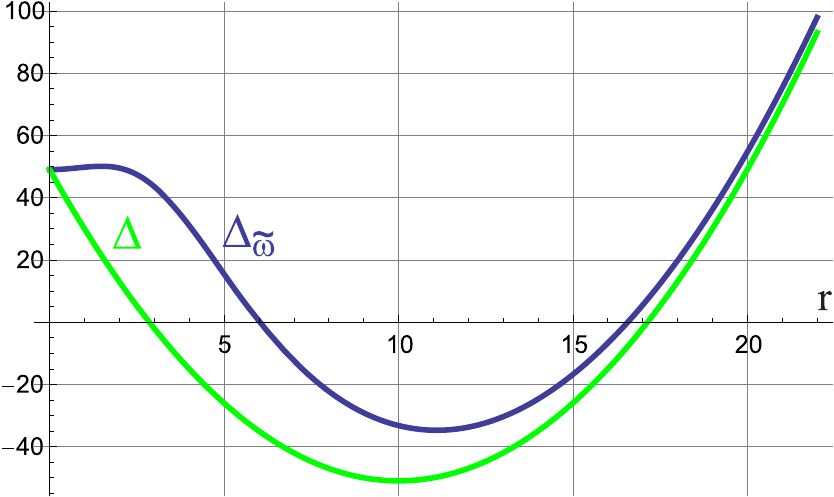}
\caption{\label{fDC&Q} A plot of $\Delta_{\tilde\omega}$ (quantum corrected case) versus $\Delta$ (classical case) for $m=$10 Planck masses and $a=7$. Note that the differences between both cases are smaller as $r$ grows.}
\end{figure}
\item As in the classical case, the number of horizons depend on the relationship between $m$ and $a$ and there can be just none, one or two horizons (see fig.\ref{fnumberh}).
\begin{figure}[h]
\includegraphics[scale=0.5]{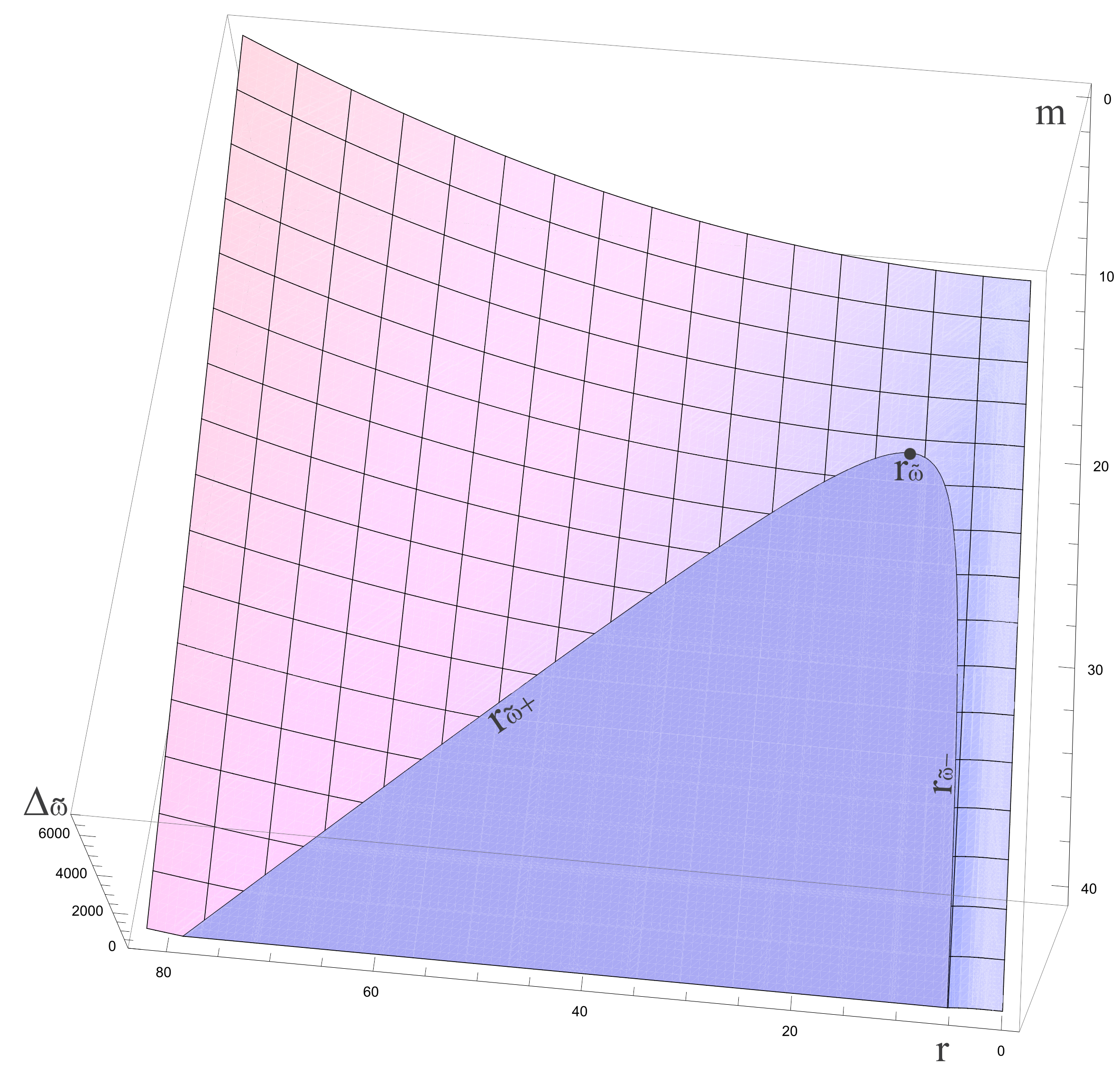}
\caption{\label{fnumberh} A plot of $\Delta_{\tilde\omega}$ (with $a=7$) as a function of the BH mass and the coordinate $r$ (with $r\geq 0$). The qualitative features are independent of the specific value chosen for $a$. The points with negative values for $\Delta_{\tilde\omega}$ have not been drawn in order to the boundary of the flat region ($\Delta_{\tilde\omega}=0$) to indicate the position of the inner ($r_{\tilde\omega-}$) and outer ($r_{\tilde\omega+}$) horizons. In this way, we observe that the number of horizons grows with the mass, starting from none for small masses, reaching the extreme case for a certain mass $m=m^*(a)$ (one horizon -denoted simply by $r_{\tilde\omega}$) and, from there, stabilizing to two: One inner and one outer horizon.}
\end{figure}

    As the figure suggests the value of $r$ for the quantum corrected inner horizon stabilizes for big enough masses satisfying $m^2\gg a^2$. In effect, in this case one can develop $G$ in the form of a series and approximately solve $\Delta_{\tilde\omega}=0$ to get
    \[
    r_{\tilde\omega -}\simeq \frac{1}{2} \sqrt{G_0 \gamma \tilde\omega+ \sqrt{G_0 \gamma \tilde\omega (8 a^2+ G_0 \gamma \tilde\omega)}},
    \]
    that in the $a=0$ case provide us with $r_{\tilde\omega -}\simeq\sqrt{\gamma\tilde\omega G_0/2}$, which is the result found in \cite{B&R} for the nonrotating case.
    Likewise, in this \textit{big mass} case one finds that the outer horizon satisfies
    \[
    r_{\tilde\omega +}\simeq G_o m+\sqrt{G_0^2 m^2-a^2}-\frac{(2+\gamma)\tilde\omega}{4 m}.
    \]
    In this way, there is a small quantum correction with respect to the classical outer horizon and, as in the non-rotating case \cite{B&R}, it affects the horizon by shrinking it.

\item The extreme case (one horizon) was obtained in the classical case whenever $a^2=m^2$. However, this is now modified by the quantum effects. For instance, it is now possible to reach the extreme case even if $a=0$ (non-rotating case \cite{B&R}). Nevertheless, since both the quantum effects and the action of the rotation help to generate an interior region with $\Delta_{\tilde\omega}>0$, the inner horizon (when it exists) tends to be bigger than the classical one and, in this way, the extreme case will be always reached for $a^2<m^2$. (See fig.\ref{fextremal}).
\begin{figure}[t]
\includegraphics[scale=1]{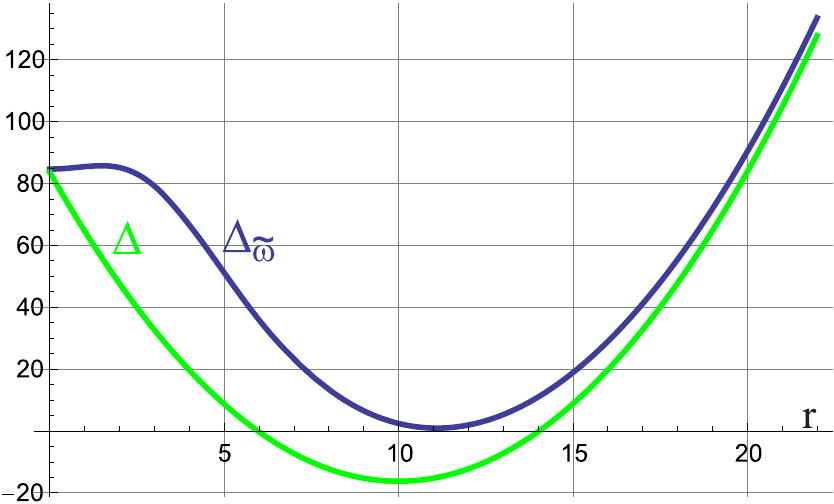}
\caption{\label{fextremal} A plot of $\Delta_{\tilde\omega}$ (quantum corrected case) versus $\Delta$ (classical case) for $m=$10 Planck masses and $a\simeq 9.151$. As can be seen, the quantum corrected case is extremal for these values, while the classical case predicts two horizons.
}
\end{figure}
\end{itemize}

\begin{figure}[ht]
\includegraphics[scale=.7]{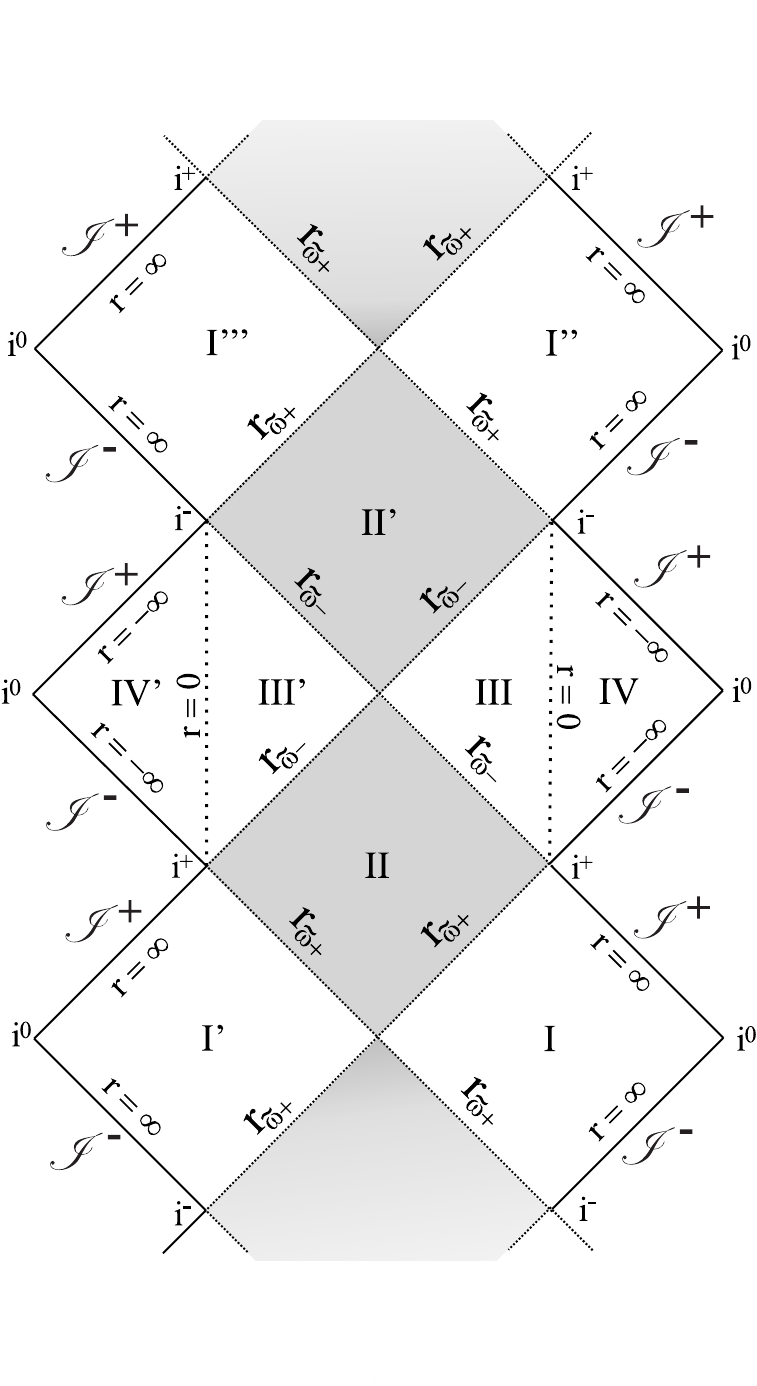}
\caption{\label{Pbiggera} Penrose diagram for a regular rotating black hole satisfying $a^2<m^{*2}$. The spacetime has been extended through $r=0$ to asymptotically flat regions with negative values for $r$ (IV or IV'). The grey regions are the regions where the coordinate $r$ is timelike. Starting from the asymptotically flat region I, one could enter region II by traversing the \emph{event horizon} $r_{\tilde\omega+}$. Region III could next be reached by traversing the \emph{Cauchy horizon} $r_{\tilde\omega-}$. Then, the asymptotically flat region IV could be reached by passing through the regular $r=0$. (Note that, since there are not singularities, unlike in Kerr's solution, the diagram is valid for all $\theta$).}
\end{figure}

\begin{figure}[ht]
\includegraphics[scale=.7]{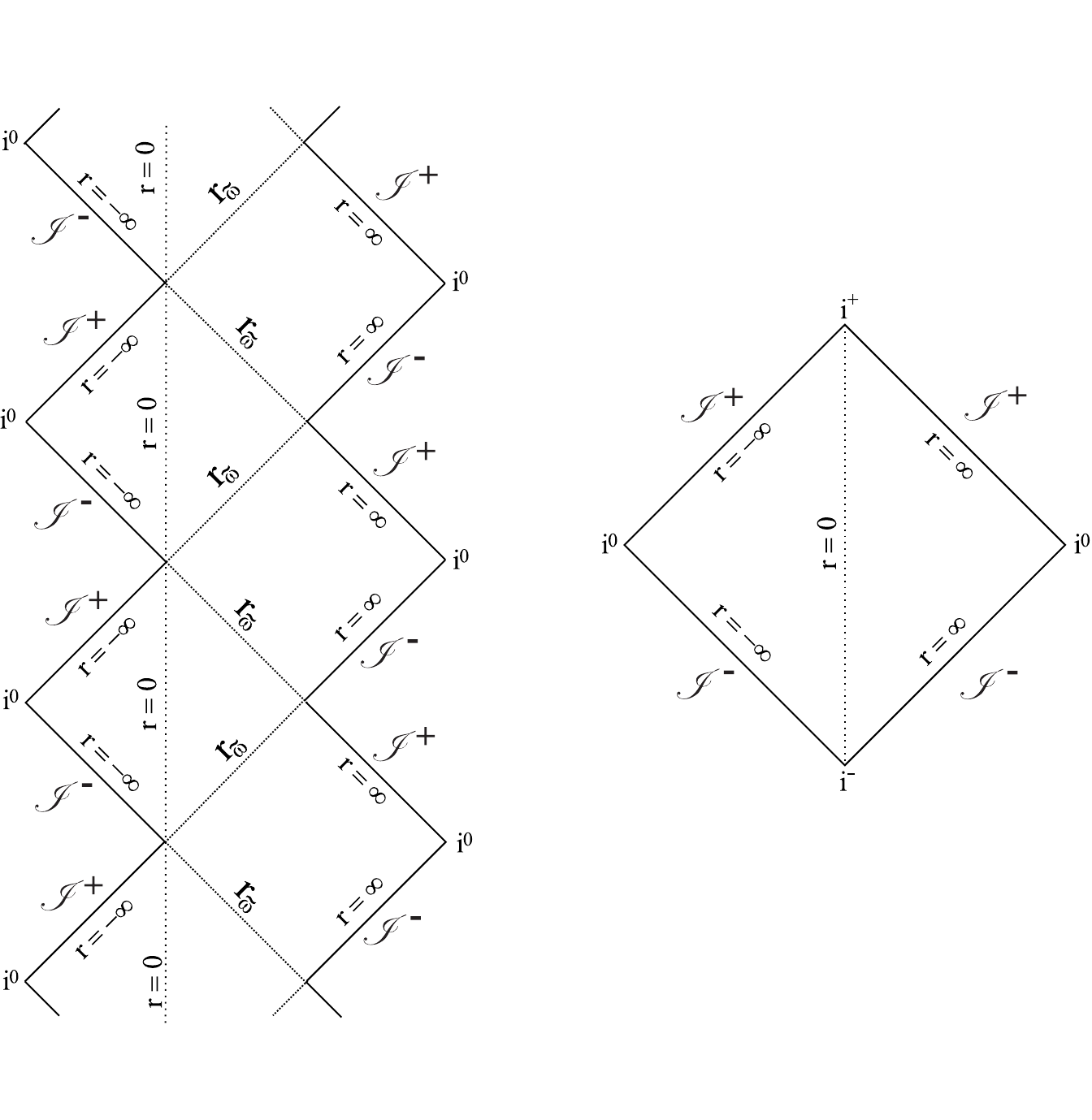}
\caption{\label{PHE} Penrose diagrams for an extreme ($a^2=m^{*2}$) regular rotating black hole (to the left) and for a hyperextreme ($a^2>m^{*2}$) regular rotating black holes (to the right). In the extreme case there is only one horizon denoted by $r_{\tilde\omega}$ in which the coordinate $r$ is lightlike. $r$ is never timelike. $r_{\tilde\omega}$ acts as both an event and a Cauchy horizon.
 In the hyperextreme case there are no horizons and $r$ is always spacelike. In both cases, the spacetime has been extended through $r=0$ to an asymptotically flat region with negative values for $r$. (Note that, again, since there are not singularities, the diagrams are valid for all $\theta$).}
\end{figure}

Let us denote by $m^*\ (=m^*(a))$ the value of the RBH mass that is needed to make a black hole of rotation parameter $a$ extreme. Then, there are three possible qualitatively different causal structures for the BH spacetime which are represented in the Penrose diagrams of figure \ref{Pbiggera} (for the $a^2<m^{*2}$ case) and of figure \ref{PHE} (for the $a^2=m^{*2}$ or \textit{extreme} case and the $a^2>m^{*2}$ or \textit{hyperextreme} case).
%

%

\section{Conclusions}\label{secCon}

The standard NJ algorithm can be used as a means of obtaining rotating spacetimes from static spherically symmetric ones. However, we have seen that, in general, its use does not provide us directly with a correct extended (through $r<0$) spacetime, neither from a mathematical point of view, nor from a physical point of view. Guided by the fact that a direct natural extension can be found in Kerr's solution, in section \ref{NJA} we have put forward a prescription in order to obtain well-behaved natural extensions for the RBH spacetimes obtained through the use of the NJ algorithm. We have seen that we could choose to extend the solution with negative values of $r$, but in order to do so, we first need to control the behaviour of the seed spacetime covered with negative values of the coordinate $r$.

We have shown this with a particular example in which the blind application of the standard algorithm provide us with an
extension (through $r<0$) of the obtained rotating solution
that turned out to be totally incorrect from a physical point of view. On the other hand, if one carries out a previous analysis of the seed spacetime covered with negative values of $r$ and uses the information in the algorithm, it provide us with a direct correct extension of the rotating solution for negative values of $r$, both from a mathematical and from a physical point of view. The obtained extension, however, is not an analytical extension since $G$ is a $C^2$ function. In fact, this is just another example in which the \textit{analytical} extension is not the correct option (see \cite{FMS} for other cases and further clarifications).

The application of the algorithm to a quantum improved solution has allowed us to obtain the extended spacetime corresponding to a regular rotating black hole that emulates the behaviour of the maximally extended Kerr solution in the regions where quantum effects are negligible -- what is in itself a very interesting result. Moreover, we have seen that the algorithm provides us with an unique running $G$ from our chosen seed solution. We have rigourously shown that the obtained spacetime does not have scalar curvature singularities and that this fact is linked to its violation of the weak energy conditions (what allows the spacetime to avoid the conditions for the existence of singularities appearing in the standard singularity theorems). In this way, while in the (classical) Kerr solution $\Sigma=0 \Leftrightarrow (r=0,\theta=\pi/2)$ defines a singular ring, in the quantum improved spacetime $\Sigma=0$ is just a \textit{regular belt}. The features of the obtained regular belt are similar to those heuristically described in \cite{B&M} and obtained for noncommutative inspired regular RBH in \cite{S&S}\cite{M&N}. However, they differ from the features found for the exact regular RBH solutions in the framework of \textit{conformal quantum gravity} \cite{BMR}, where the spacetime is inextendible beyond ``$r=0$'' and the curvature invariants are continuous.

We have seen that there are three qualitatively different cases for the obtained regular rotating black hole according to the relationship between $m$ and $a$ what, in fact, is similar to the situation found in Kerr's case. In particular, we have seen that the number of horizons and the corresponding causal structures in the classical and quantum-improved cases are strongly related. However, the position of the horizons is modified due to the repulsive character of the quantum improvements. In this way, the inner horizon is bigger than the classical inner horizon, while the outer horizon shrinks with respect to the classical one. Related to this effect, we get that the extreme case is obtained for smaller rotations than in the classical case when quantum improvements in the RBH spacetime are considered (i.e., it is obtained for $a^2<m^2$).

It must be taken into account that the reliability of the QEG approach used to obtain the seed \textit{improved Schwarzschild solution} \cite{B&R} is questionable in the planckian regime, so that the \emph{regular belt} is just suggested by the approach, but can not be guaranteed. Indeed, only a still nonexistent full Quantum Gravity Theory could provide us with the exact description in the planckian regime.

Finally, it is necessary to remark that, in general, there are other possible extensions (`beyond $r=0$') for rotating black hole spacetimes, apart from the `$r<0$' extension discussed here. In each case, every possible extension has its own mathematical and/or physical pros and cons. A full analysis of the different extensions for general rotating black hole spacetimes will be the subject of a future work \cite{future}.

\section*{Acknowledgements}
R Torres acknowledges the financial support of the Ministerio de Econom\'{\i}a y Competitividad (Spain), projects MTM2014-54855-P.

\appendix
\section{Running G for $r<0$}
In the introduction we stated that the Functional Renormalization Group Equation leads to a running $G$ with the form \cite{B&R}\cite{Reuter}
\[
G(k)=\frac{G_0}{1+\omega G_0 k^2}.
\]
Then, one converts the energy scale dependence into a position dependence, what can be written as
\[
k(P)=\frac{\xi}{d(P)},
\]
where $\xi$ is a numerical constant to be fixed and $d(P)$ is the \textit{distance scale} that provides the relevant cutoff when a test particle is located at a point $P$.
Finally, if the distance scale $d(P)$ form the point $P$ to the center of the black hole must be diffeomorphism invariant then one could write
\[
d(P)=\int_{\mathcal C} \sqrt{|ds^2|},
\]
where $\mathcal C$ is a curve joining the points. In the case of a spherically symmetric BH the symmetry imposes that $d=d(r)$, however one still has to find an expression for the function, what requires considering the different possibilities for $\mathcal C$.

So far, we have been following the procedure described in \cite{B&R}. We will still do it, with the sole difference that now we want to consider $r<0$. It is straightforward to see that Schwarzschild's solution has no horizons for $r<0$ (or, equivalently, for negative masses) and that the coordinates $r$ and $t$ remain spacelike and timelike, respectively, for all $r<0$, what in fact makes the computations easier than in the $r>0$ case. First, let us consider the \textit{radial} curve $\mathcal C_1$: $r=\lambda, t=t_0, \theta=\theta_0,\phi=\phi_0$. We have for all $r\leq 0$
\begin{eqnarray}\label{exact}
d_1(r)=\int_r^0 \left(1-\frac{2 G_0 m,}{r}\right)^{-1/2}\ dr=\nonumber\\
=\sqrt{r (r-2 G_0 m)}-2 G_0 m \tanh^{-1} \sqrt{\frac{r}{r-2 G_0 m}}.
\end{eqnarray}
(Note that $d(r<0)>0$). The behaviour of this function for $|r|\ll G_0 m$ is
\begin{equation}\label{nsb}
d_1(r)\simeq \frac{2}{3} \frac{1}{\sqrt{2 G_0 m}} |r|^{3/2}
\end{equation}
while for $|r|\gg G_0 m$
\begin{equation}\label{fsb}
d_1(r)\simeq |r|.
\end{equation}
This is exactly the behaviour obtained for $r>0$ in \cite{B&R}, with the only difference that we have to add a modulus ($||$) to our negative $r$. Likewise, following \cite{B&R}, it is easy to see that other curves provide the same behaviour (\ref{nsb}) for $|r|\ll G_0 m$, while for $|r|\gg G_0 m$ the behaviour (\ref{fsb}) provides the largest momentum scale and, therefore, the actual cutoff. In this way, even if one cannot assert that (\ref{exact}) provide us with the exact behaviour of the distance scale, one concludes that the correct qualitative behaviour should interpolate between $|r|^{3/2}$ and $|r|$, what suggest to use in concrete computations the \textit{interpolating distance scale}
\[
d(r\leq 0)=\left(\frac{r^3}{r-\gamma G_0 m} \right)^{1/2}
\]
with $\gamma=9/2$. Now, using $k(r)=\xi/d(r)$ and the expression for the running $G$
\[
G(r\leq 0)=\frac{G_0 r^3}{r^3+\tilde\omega G_0 (r-\gamma G_0 m)},
\]
where $\tilde\omega\equiv\omega \xi^2$.
Therefore, as stated in (\ref{Gallr}), the behaviour for $r\geq 0$ (\ref{GR}) and the just found behaviour for $r\leq 0$ can be combined in a running $G$ for all $r$ as
\[
G(r)=\frac{G_0 |r|^3}{|r|^3+\tilde\omega G_0 (|r|+\gamma G_0 m)}.
\]

%

\end{document}